\documentclass{article}
\usepackage{amsfonts}
\usepackage{spconf,amsmath,graphicx,epsfig}

\title{Tensor-to-Vector Regression for Multi-channel Speech Enhancement based on Tensor-Train Network}
\name{Jun Qi$^{1*}$, Hu Hu$^{1*}$, \thanks{* Refers to an equal contribution.}
 Yannan Wang$^{3}$, Chao-Han Huck Yang$^{1}$, Sabato Marco Siniscalchi$^{1,2}$, Chin-Hui Lee$^{1}$ }
\address{$^1$Electrical and Computer Engineering, Georgia Institute of Technology, Atlanta, GA, USA \\
$^2$Computer Engineering School, University of Enna, Italy \\
$^3$Tencent Media Lab, Tencent Corporation, Shenzhen, Guangdong, China}
\begin{document}
\ninept
\maketitle
\begin{abstract}
We propose a tensor-to-vector regression approach to multi-channel speech enhancement in order to address the issue of input size explosion and hidden-layer size expansion. The key idea is to cast the conventional deep neural network (DNN) based vector-to-vector regression formulation under a tensor-train network (TTN) framework.  TTN is a recently emerged solution for compact representation of deep models with fully connected hidden layers. Thus TTN maintains DNN's expressive power yet involves a much smaller amount of trainable parameters. Furthermore, TTN can handle a multi-dimensional tensor input by design, which exactly matches the desired setting in multi-channel speech enhancement.  We first provide a theoretical extension from DNN to TTN based regression. Next, we show that TTN can attain speech enhancement quality comparable with that for DNN but with much fewer parameters, e.g., a reduction from 27 million to only 5 million parameters is observed in a single-channel scenario.  TTN  also improves PESQ over DNN from 2.86 to 2.96 by slightly increasing the number of trainable parameters. Finally, in 8-channel conditions, a PESQ of 3.12 is achieved using 20 million parameters for TTN, whereas a DNN with 68 million parameters can only attain a PESQ of 3.06. Code is available online\footnote{https://github.com/uwjunqi/Tensor-Train-Neural-Network}. 
\end{abstract}
\begin{keywords}
Tensor-Train network, speech enhancement, deep neural network, tensor-to-vector regression
\end{keywords}

\section{Introduction}
\label{sec1}
Deep neural network (DNN) based speech enhancement \cite{xu2015regression} has demonstrated state-of-the-art performances in a single-channel setting. It has also been extended to multi-channel speech enhancement with similar high-quality enhanced speech \cite{wang2018two}. A recent overview can be found in \cite{dwang2018}. In essence, the process can be abstracted as a vector-to-vector regression based on deep architectures with a DNN aiming at learning a functional relationship $f: \mathbb{Y} \rightarrow \mathbb{X}$ such that input noisy speech $ y \in \mathbb{Y}$ can be mapped to  corresponding clean speech $ x \in \mathbb{X}$. Several variants of deep learning structures have also been attempted, e.g., recurrent neural networks (RNNs) with long short term memory (LSTM) gates were employed in ~\cite{weninger2015speech, zhao2018convolutional}. A deep bidirectional RNN with LSTM gates was instead used in~\cite{sun2015voice}. Moreover, a generative adversarial network (GAN) was used in~\cite{pascual2017segan}.

\begin{figure}[t]
\centerline{\epsfig{figure=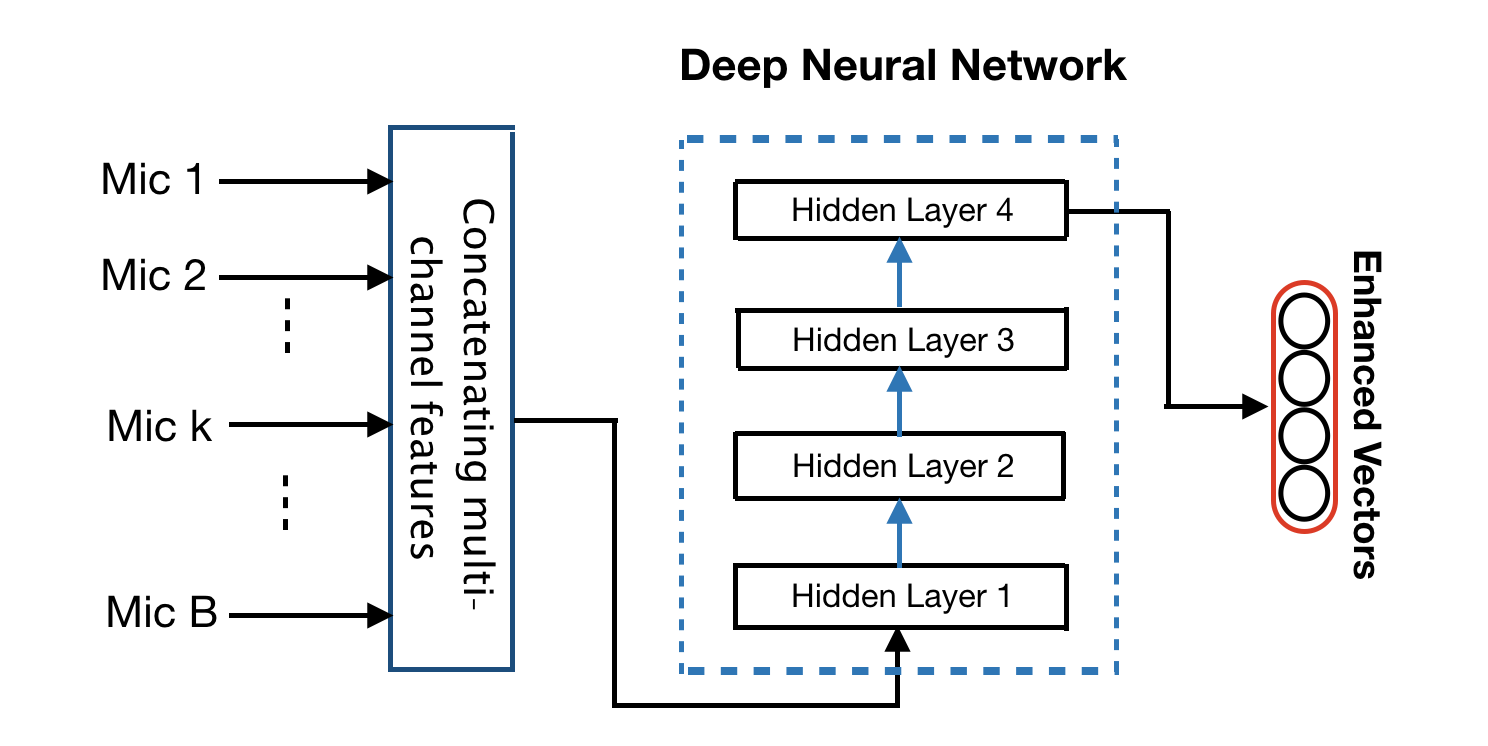, width=85mm}}
\caption{{\it Conventional multi-channel DNN-based vector-to-vector regression for speech enhancement. }}
\label{fig:fig1}
\end{figure}

Spatial information can complement spectral information for improved speech enhancement leveraging multi-channel information in more complex scenarios, where the sources of distortions include ambient noises, room reverberation and interfering speakers, as discussed in \cite{vantrees2004, benesty2008, Xiao2016a, wang2018}, for example. However, the DNN-based vector-to-vector regression, which is the focus in this work, mainly aims at single-channel speech enhancement and is not simply generalized to multi-channel speech enhancement. As shown in Figure~\ref{fig:fig1}, a traditional approach to dealing with an array of microphones is exploited spatial information at the input level by concatenating speech vectors from multiple microphones into a single high dimensional vector, e.g.,~\cite{wu2017multichannel, wang2018two}. Thus the vector-to-vector regression approach can still be employed for speech enhancement by appending multi-channel feature vectors together into a high-dimensional vector and mapping it to a vector extracted from the reference vector. Such a simple solution, unfortunately, clashes with our theoretical analysis outlined in~\cite{qi2019theory}, which suggests that the width of each hidden layer in a DNN needs to be greater than the input dimension plus two so that the expressive power of DNN-based vector-to-vector regression can be guaranteed. Moreover, high dimensional input vectors result in wider hidden non-linear layers, which in turn implies the need for huge computational resources and storage cost. Several proposals were put forth by different groups to overcome such an issue, for example, sparseness in DNN was explored in~\cite{yu2012} to reduce the model size; however, the optimum implementation heavily depends on the specific hardware architecture. In \cite{xue2013}, a singular value decomposition approach was devised, which is performed in a post-processing phase, that is, a DNN with a huge amount of parameters is first trained, and parameter reduction is then applied.

\begin{figure}[t]
\centerline{\epsfig{figure=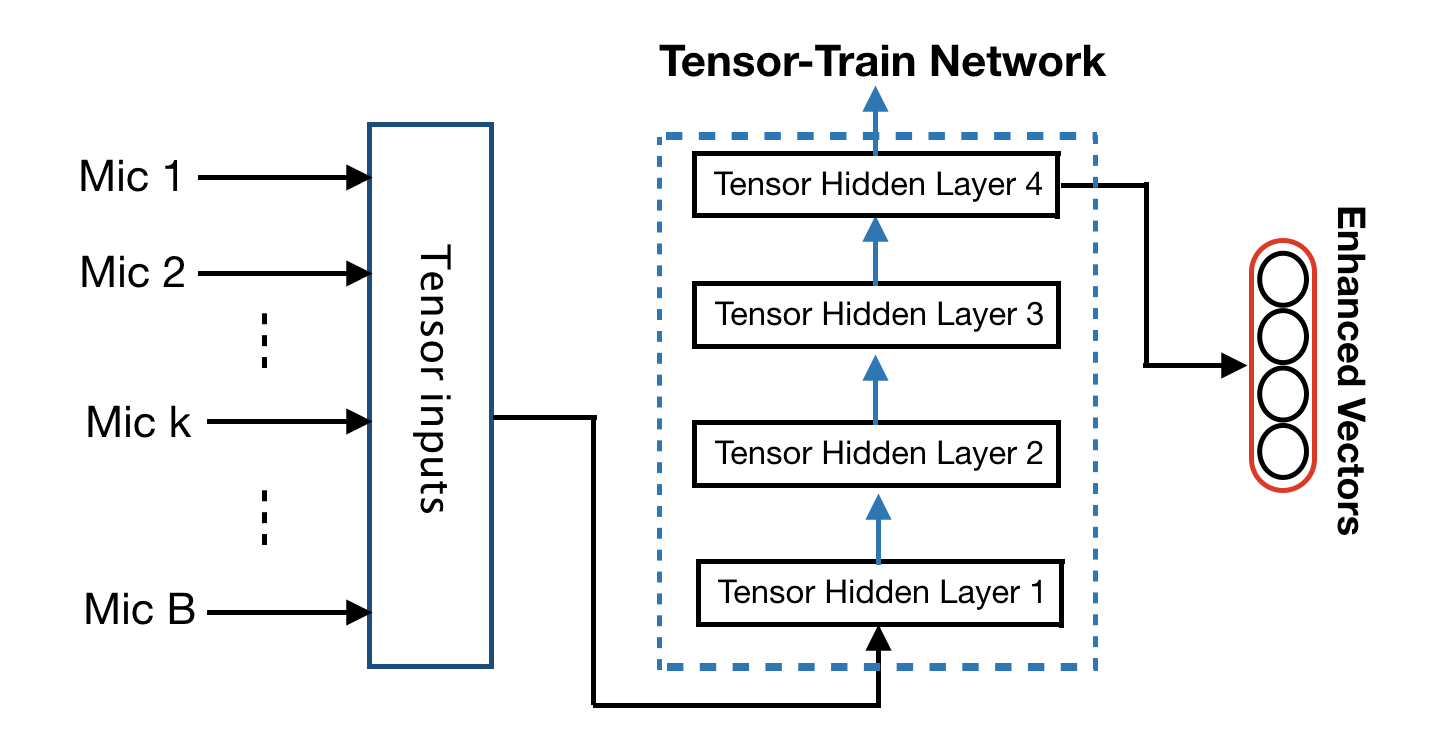, width=85mm}}
\caption{{\it A TTN-based speech enhancement system.}}
\label{fig:fig2}
\end{figure}

In this work, we proposed to address the multi-channel speech enhancement problem within a more principled parameter reduction framework, namely the Tensor-Train Network (TTN) \cite{novikov2015tensorizing}, which does not require a multi-stage approach. A Tensor-Train refers to a compact tensor representation of the fully-connected hidden layers in a DNN~\cite{novikov2015tensorizing}. In doing so, we put forth a tensor-to-vector regression approach that can handle multi-channel information and meanwhile addresses the issue of input size explosion and hidden-layer size expansion. As shown in Figure~\ref{fig:fig2}, the feature vectors of $B$ microphones are transformed into a tensor representation,
such that the DNN-based vector-to-vector regression can be reshaped to a TTN based tensor-to-vector mapping. We show how the theorem on expressive power of the traditional DNN-based vector-to-vector \cite{qi2019theory} can be extended to be valid in the TTN framework. Next, we evaluate our approach with both single- and multi-microphone experimental setups, where the speech signal is  exposed to multiple sources of distortions, including ambient noises, room reverberation and interfering speakers.  We show that TTNs can attain speech enhancement quality comparable with DNNs, yet  a 81\% reduction in the size of parameters is observed, in the single-channel scenario. In 8-channel conditions, TTN  delivers a PESQ of 3.12 using  20 million parameters. In contrast, a DNN-based multi-channel configuration needs up to 68 million parameters and attain a PESQ of 3.06. We should also remark that Tensor-Train decomposition can be applied to CNNs, and RNNs; however,  we focus on  DNNs in this initial investigation.

The remainder of the paper is organized as follows: Section \ref{sec:sec2} introduces TTN mathematical underpinnings and key properties. In Section \ref{sec:exp}, the experimental environment is first described, and then the experimental results are presented and discussed. Section~\ref{sec:conclusion} concludes our work.

\section{Tensor-Train Network}
\label{sec:sec2}
TTN relies on a tensor-train decomposition~\cite{oseledets2011tensor} which can be described mathematically as follows: given a vector of ranks $\textbf{r} = \{ r_{1}, r_{2}, ..., r_{K+1} \}$, tensor-train decomposes a tensor $W \in \mathbb{R}^{(m_{1}n_{1}) \times (m_{2}n_{2})\times \cdot\cdot\cdot \times (m_{K}n_{K})},  \forall i\in \{1, ..., K\}, m_{i} \in \mathbb{R}^{+}, n_{i}\in \mathbb{R}^{+}$ into a multiplication of core tensors according to Eq.~(\ref{eq:ttf}), where for the given ranks $r_{k}$ and $r_{k+1}$, the $k$-th core tensor $C^{[k]}(r_{k}, i_{k}, j_{k}, r_{k+1}) \in \mathbb{R}^{m_{k} \times n_{k}}$ in which $i_{k} \in \{1, 2, ..., m_{K}\}$ and $j_{k} \in \{1, 2, ..., n_{K}\}$. Besides, $r_{1}$ and $r_{K+1}$ are fixed to $1$.
\begin{equation}
\label{eq:ttf}
W((i_{1}, j_{1}), (i_{2}, j_{2}), ..., (i_{K}, j_{K})) = \prod\limits_{k=1}^{K} C^{[k]}(r_{k}, i_{k}, j_{k}, r_{k+1})
\end{equation}

TTN is generated by applying the Tensor-Train decomposition to the hidden layers of DNN. The key benefit from TTN is to significantly reduce the number of parameters of a feed-forward DNN with fully-connected hidden layers. It is because TTN only stores the TT-format of DNN, i.e., the set of low-rank core tensors $\{C_{k}\}_{k=1}^{K}$ of the size $\sum_{k=1}^{K}m_{k}n_{k} r_{k} r_{k+1}$, which can approximately reconstruct the original DNN. In contrast, the memory storage of original DNN requires the size of $\prod_{k=1}^{K} m_{k} n_{k}$, which is much larger than $\sum_{k=1}^{K}m_{k} n_{k} r_{k} r_{k+1}$. Moreover, instead of decomposing a well-trained DNN into a TTN, core tensors can be randomly initialized and iteratively trained from scratch. Similar to the DNN training procedures, the optimization methods based on variants of stochastic gradient descent (SGD) can ensure converged solutions. Besides, the work~\cite{novikov2015tensorizing} shows that the running complexities of DNN and the corresponding TTN are in the same order scale.

\section{Tensor-to-Vector Regression}
\label{sec:sec3}
\begin{figure}[htbp]
\centerline{\epsfig{figure=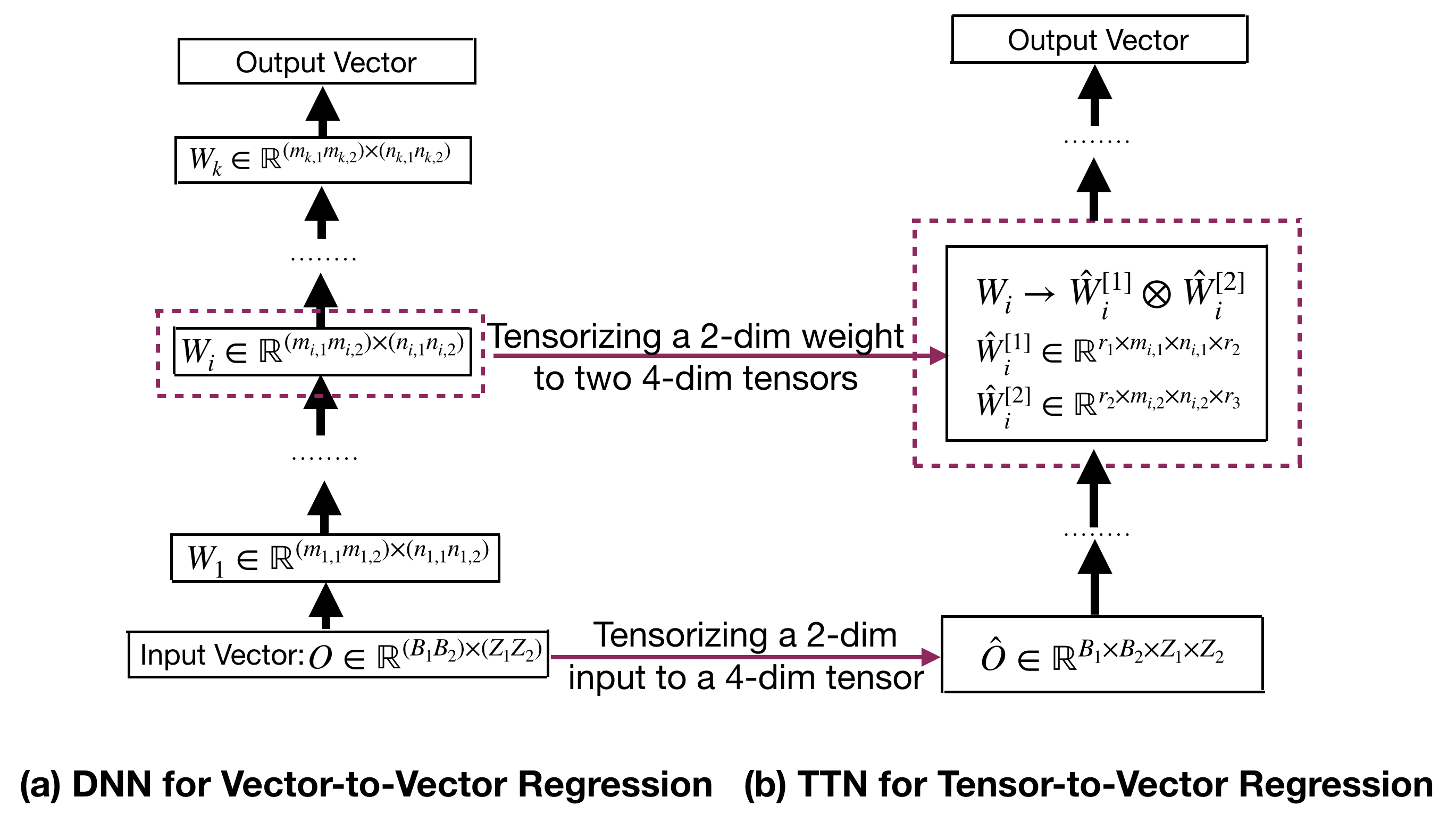, width=85mm}}
\caption{{\it Transforming a DNN-based vector-to-vector regression into a TTN-based tensor-to-vector regression.}}
\label{fig:fig3}
\end{figure}

The TTN framework offers a natural way to convert a speech enhancement system from vector-to-vector to tensor-to-vector configuration because multi-dimensional inputs can be directly fed into the TNN avoiding concatenating speech vectors inputs into a single long vector. Figure~\ref{fig:fig3} demonstrates a typical example of casting DNN-based vector-to-vector regression into a TTN-based tensor-to-vector regression. More specifically, given a list of ranks $\{r_{1}, r_{2}, r_{3}\}$, the DNN input vector $O \in \mathbb{R}^{(B_{1}B_{2})\times (Z_{1}Z_{2})}$ is decomposed into a 4th order input tensors $\hat{O} \in \mathbb{R}^{B_{1} \times B_{2} \times Z_{1} \times Z_{2}}$ before being used in the TTN-based configuration. Accordingly, the $i$-th weight matrix,  $W_{i}$,  in the DNN is decomposed into a tensor product of two tensors $\hat{W}_{i}^{[1]} \in \mathbb{R}^{r_{1} \times m_{i,1}\times n_{i,1}\times r_{2}}$ and $\hat{W}_{i}^{[2]} \in \mathbb{R}^{r_{2} \times m_{i,2}\times n_{i,2}\times r_{3}}$. The corresponding hidden layers with core tensors represent a basic architecture of TTN for tensor-to-vector regression, and the output vectors of two architectures are set to be equal.

Tensor-to-vector regression based on TTN can substantially reduce the number of model parameters. Besides, we can prove that the TTN-based tensor-to-vector regression can maintain the representation power of the traditional DNN-based vector-to-vector. Our theoretical work~\cite{qi2019theory} on the representation power of DNN-based vector-to-vector regression suggests a tight upper bound to a numerically estimate of the maximum mean squared estimation (MSE) loss. That MSE upper bound explicitly links the depth of ReLU-based hidden layers to the expressive capability of DNN models. More specifically, with input and out dimensions of $d$ and $q$, respectively for a vector-to-vector regression target function $\hat{f}: \mathbb{R}^{d} \rightarrow \mathbb{R}^{q}$, there exists a DNN $f_{DNN}$ with $K (K \ge 2)$ ReLU based hidden layers, where the width of each hidden layer is at least $(d+2)$ and the top hidden layer has $n_{K} (n_{K} \ge d + 2)$ units. Then, we can derive the equality of (b) in Eq.~(\ref{eq:bound}).

\begin{equation}
\label{eq:bound}
|| \hat{f} - f_{TTN} ||_{1} \overset{(a)}{\le} || \hat{f} - f_{DNN} ||_{1} \overset{(b)}{=} O(\frac{q}{(n_{K} + K - 1)^{2}})
\end{equation}

We are currently developing new theories on TTN-based tensor-to-vector mapping, which demonstrate that a TTN associated with the tensor decomposition of a DNN can keep the expressive power of DNN-based vector-to-vector regression. Indeed, we can mathematically find a TTN $f_{TTN}$ representation such that the inequality of (a) in Eq.~(\ref{eq:bound}) is valid. We avoid giving the proof of our new results here, but the experimental evidence given in the next section on the speech enhancement task supports our theoretical results in Eq.~(\ref{eq:bound}).


The inequality in Eq.~(\ref{eq:bound}) is ensured to be valid under consistent training and testing environmental conditions. It basically states that a TTN has more expressive power than a DNN with a similar complexity. The detailed proof is shown in our another paper \cite{junciss}, and the effect is demonstrated in the experimental results in Tables~\ref{tab:single} and~\ref{tab:multi} to be provided later. 



\section{Experimental Setup \& Results}
\label{sec:exp}

\subsection{Data Preparation}
The proposed TTN-based models are evaluated on simulated data from WSJ0, which contains additive noise, interfering speakers, and reverberation. The dataset is created by corrupting the WSJ0 corpus~\cite{paul1992design} with OSU-$100$-noise \cite{osu100} data, allowing us to obtain $30$ hours of training material, and $5$ hours of testing data. When simulating the noisy data, each waveform is mixed with one kind of background noise from the noise set. The target and additional interfering speech with their corresponding RIRs are convolved to generate the final waveform. In particular, our training and testing datasets are created from different noisy utterances of various speakers. As for the training dataset, a $5$-minute clean speech from each of the targeted speakers is randomly mixed with $73$ interfering speakers and $90$ types of additive noise. Each target an-echoic speech is generated by convolving clean speech with the direct path response between the target speakers and the reference channel. For the testing dataset, another $5$-minute unseen speech of the target speakers is mixed with $10$ unseen interfering speakers and $10$ types of unseen noise.
The signal-to-interfered-noise-ratio (SINR) level of each utterance is set as follows: when SINR is $5$dB, the signal-to-noise-ratio (SNR) is configured to $10$dB or $15$dB; when SINR is $10$dB, the SNR is set to $15$dB; when SINR is $15$dB, the SNR is increased to $20$dB. The proportion of each SINR level is set equally. Besides, some utterances of SINR $30$dB are included in the training set to cover some very high SINR conditions.

To simulate reverberated speech, a reverberated acoustic environment is built: A microphone array of $8$-circular channel microphone is arranged in a room of size $6.5m \times 5.5m \times 3m$ in terms of length-width-height. As to the single-channel scenario, the microphone is put at the center of the array. To avoid unnecessary combinations of multiple interference, we deliberately constrain the conditions that the microphone array only aims at one target speaker, and it only received one kind of additive noise. Specifically, a horizontal distance of a target speaker to the center of the microphone array is strictly fixed to $3m$. Besides, we set both the target speaker and the interfering speaker keeping the same distance to the microphone array, and the angle of them is configured as $40^{\circ}$. Before we build the training and testing sets, an improved image-source method (ISM) \cite{data-ism} is used to generate RIRs of reverberation time (RT$60$) (from $0.2s$ to $0.3s$) and the corresponding direct path response for each microphone channel. For both training and testing datasets, the setting of RIRs is fixed to the same conditions, such as the room size, RT$60$, and all of the distances and directions.

\subsection{Experiment Setting}
In our experiments, $257$-dimensional normalized log-power spectral (LPS) feature~\cite{hou2007saliency, qi2013auditory, feature2} is taken as the inputs to the DNNs. The LPS features are generated by computing $512$ points Fourier transform on a speech segment of $32$ milliseconds. For $B$-channel data, all channel inputs are concatenated together for the model training. For each input frame, the adjacent context of size $M$, is combined with the current frame. Hence, the size of the input for DNN is $257 \times (2M+1) \times B$. For TTN, we ignore the first dimension of the input LPS features, which is the direct-current component. Thus, the input size for TTN is $256 \times (2M+1) \times B$. After regression, the first dimension of input is concatenated back to the $256$-dimension output without any change. The clean speech features of the first channel are assigned to the top layers of DNN and TTN, as the reference during the training stage.

Our baseline DNN model has $6$ hidden layers, and each hidden layer is composed of $2048$ neurons. The ReLU activation function is used for all hidden layers. A linear function is employed in the top hidden layer. As for the TTN, each hidden dense layer is decomposed and replaced with a Tensor-Train format, as shown in Figure~\ref{fig:fig2}. Both the DNN and TTN are trained from scratch based on the standard back-propagation algorithm~\cite{hirose1991back}, and both models adopt the same training configuration. During the training phase, Adam optimizer is adopted, and the initial learning rate is set to $0.0002$. The mean square error (MSE) \cite{mse, qi2013subspace} is used as the optimization object. The context window size at the input layer is set to $5$ for all models, in which the current frame is concatenated with the previous $5$ frames and following $5$ frames (from the same channel). For both the DNN and TTN models, TensorFlow~\cite{tensorflow} is used in all of our experiments. 

Perceptual evaluation of speech quality (PESQ) \cite{rix2001perceptual} is used as our evaluation criterion. The PESQ score, which ranges from $-0.5$ to $4.5$, is calculated by comparing the enhanced speech with the clean one. A higher PESQ score corresponds to a higher quality of the speech perception.

\subsection{Single-channel Speech Enhancement Results}
The TTN framework is first evaluated on the single-channel speech enhancement task. In Table~\ref{tab:single}, we can see that a $6$-layer DNN model, which is taken as a baseline system, achieves a PESQ score of $2.86$ with $27$ million parameters. A TTN architecture is generated by applying tensor-train decomposition to the DNN baseline model. Each weight matrix of DNN is decomposed to two four-dimension tensors using tensor decomposition. For example, a weight matrix of the size $2048 \times 2048$ can be decomposed to two tensors with the size of $1 \times 32 \times 32 \times 4$ and $4 \times 64 \times 64 \times 1$. The TTN core tensors are randomly initialized and then trained from scratch using the Adam optimizer \cite{diederik2014}. 

As discussed in Sections~\ref{sec:sec2} and~\ref{sec:sec3}, the forth row in Table~\ref{tab:single} shows that a substantial parameter reduction when the Tensor-Train format is employed. A drop in the PESQ value, from $2.86$ (DNN) to $2.66$ (TNN), is observed because of the parameter reduction. Nonetheless, TTNs consistently delivers better and better speech enhancement results as its number of parameters increases.  As shown in the seventh row in Table~\ref{tab:single}, the TTN model with $5$ million parameters can achieve nearly the same PESQ scores as the DNN baseline model ($2.84$ vs $2.86$). However, TTN model uses only  $18\%$ of the amount of parameters in the DNN, which suggests that TTN can significantly reduce the number of parameters while keeping the baseline performance. Furthermore, if we further increase the parameter number of the TTN model, we can even obtain a better TTN model achieving a $0.1$ absolute PESQ improvement using only $74\%$ of the DNN  parameters, namely $20$ million (see last row in the table).

\begin{table}[ht]
    \caption{PESQ results for single-channel speech enhancement.}
    \vspace{0.2cm}
    \centering
    \begin{tabular}{l|c|c|c}
    \hline
    \hline
    Model  & Channel \#  & Parameter \#& PESQ \\
    \hline
    \hline
    DNN    & 1                        & 27M          & 2.86 \\
    DNN-SVD    & 1                        & 5M     & 2.82 \\
    DNN-SVD    & 1                        & 20M      & 2.84  \\
    \hline
    TTN & 1               & 0.6M         & 2.66 \\
    TTN & 1               & 0.7M       & 2.71 \\
    TTN & 1               & 2.6M        & 2.78 \\
    TTN & 1               & 5M         & 2.84 \\
    TTN & 1               & 20M        & 2.96 \\
    \hline\hline
    \end{tabular}
    \label{tab:single}
\end{table}

To better appreciate the TTN technique, we have implemented the DNN-SVD method proposed in \cite{xue2013}, which is a widely used post-processing compression technique for deep models. Each weight matrix in the trained DNN is first decomposed into two smaller matrix using SVD. The reduced size DNN is then fine-tuned with back-propagation. The DNN-SVD method can allow us to reduce the DNN size to $5$ and $20$ million parameters, as shown in the second and third rows in Table \ref{tab:single}, respectively. Nonetheless, the PESQ drops to $2.82$ with $5$ million parameters. The PESQ attains a value of $2.84$ when $20$ million parameters are kept after DNN-SVD, which is however still lower than the PESQ attained by the original DNN. Most importantly, our TNN approach with $20$ million parameters attains a PESQ of of $2.96$, which not only compares favourably with the DNN-SVD solution with the same amount of parameters, but it represents an improvement over the DNN baseline approach, as already mentioned.

\subsection{Multi-channel Speech Enhancement Results}
\label{multi}
In this section, our investigation focuses on the microphone array scenario. The DNN-based multi-channel speech enhancement solutions realized in Figure \ref{fig:fig1}  handles information from different mics at the input layer, which calls for wider input vectors and thereby an overall larger model size. By comparing the DNN baseline configuration in the first row in Tables~\ref{tab:single} and ~\ref{tab:multi}, the parameter number goes from $27$ million for the $1$-channel case to $33$ million for $2$-channel case. In the $8$-channel case, the number of DNN parameters goes up to $68$ million, as shown in the third row in Table ~\ref{tab:multi}. In multi-channel speech enhancement, the control of  the size of the neural architecture therefore becomes even a more important aspect that cannot be overlooked.

\begin{table}[t]
    \caption{PESQ Results of Multi-channel Speech Enhancement}
    \vspace{0.2cm}
    \centering
\begin{tabular}{l|c|c|c}
\hline
\hline
Model  & Channel \# & Parameter \# & PESQ \\
\hline
\hline
DNN    & 2                  & 33M     & 3.00 \\
DNN-SVD & 2                 & 5M      & 2.93  \\
DNN    & 8                  & 68M    & 3.06 \\
 DNN-SVD    & 8              & 5M    &  3.01 \\
\hline
TTN & 2        & 0.6M   & 2.75 \\
TTN & 2        & 5M      & 2.96 \\
TTN & 8        & 0.6M  & 2.83 \\
TTN & 8        & 5M    & 3.06 \\
TTN & 8        & 20M  & 3.12 \\
\hline\hline
\end{tabular}
    \label{tab:multi}
\end{table}

In the $2$-channel configuration, TNN parameters can be reduced to $0.6$ million, but the PESQ would go down to $2.75$, as shown in the fourth row in Table~\ref{tab:multi}. Nevertheless, we can observe, by comparing the first and sixth row in  Table~\ref{tab:multi}, that the TTN-based speech enhancement model can attain nearly the same performance of the DNN-based counterpart ($3.00$ vs. $2.96$), while squeezing the number of parameters from $33$ million to $5$ million. As in the single-channel case, DNN-SVD could be applied as a post-processing approach for model compression, and it can achieve a PESQ of $2.93$ with $5$ million parameters, as shown in the fifth row in Table~\ref{tab:multi}. In the $8$-channel configuration, the TNN model size can be significantly reduced to $7\%$ of the corresponding DNN size, namely from $68$ million to $5$ million, as shown in the eighth row in Table \ref{tab:multi}, while keeping the PESQ equal to $3.06$. When compared with the result of DNN-SVD method \cite{xue2013} reported in the fourth row in Table~\ref{tab:multi}, although both DNN-SVD and TTN can reduce the model parameters to $5$ million, the PESQ will decrease from $3.06$ to $3.01$ with the DNN-SVD method; whereas the TTN approach keeps the model performance. Such an empirical result demonstrates the advantages in using a TTN-based tensor-to-vector regression approach to multi-channel speech enhancement and connects our theoretical analysis in Section~\ref{sec:sec3}. Furthermore, better speech enhancement results can be attained by increasing the TTN model size. Indeed, a TTN model with $20$ million parameters, compared to the $68$ million of the DNN baseline system, can achieve a PESQ of $3.12$, as shown in the last row in Table \ref{tab:multi}, which corresponds to a $0.6$ absolute improvement over the DNN baseline solution.

Since complex noisy and reverberated conditions are involved, the theory on the expressive power as shown in Eq.~(\ref{eq:bound}), is not strictly valid in our testing cases. But our improved theorems, which are to be published elsewhere and consider the generalization capability of models, can better correspond to our experimental results in Table~\ref{tab:single} and~\ref{tab:multi}.

\vspace{-0.3cm}
\section{Conclusions}
\label{sec:conclusion}
This work is concerned with  multi-channel speech enhancement leveraging upon TTN for tensor-to-vector regression. First, we describe the tensor-train decomposition, and its application to a DNN structure. Our theoretical results justify that the TTN-based tensor-to-vector function allows fewer parameters to realize a higher expressive and generalization power of DNN-based vector-to-vector regression. Then, we set up both single-channel and multi-channel speech enhancement systems based on TTN to verify our claims. When compared with DNN-based speech enhancement systems,  experimental evidence  shows that  tensor-to-vector regression based on TTN compares favorably with the DNN-based baseline counterpart in terms of PESQ, but it requires much fewer parameters. Furthermore, increasing number in the TTN-based configuration, we can achieve better enhancement qualities in terms of PESQ, while amount of of parameters used in the TTN configuration is much less than the DNN counterpart. The related empirical results verify our theories and exhibit a potential usage of tensor-train decomposition in more advanced deep learning models. Our future work will apply TTN decomposition to other deep models, such as recursive neural networks (RNNs) and convolution neural networks (CNNs).

\clearpage
\bibliographystyle{IEEEbib}
\bibliography{refs}

\end{document}